\documentclass{article}
\usepackage{spconf,graphicx}
\usepackage{times} 
\usepackage{latexsym} 
\usepackage[T1]{fontenc} 
\usepackage[utf8]{inputenc} 
\usepackage[colorlinks=true, linkcolor=blue, citecolor=blue, urlcolor=blue]{hyperref}
\usepackage{pifont}
\usepackage{amsmath} 
\usepackage{amssymb} 
\usepackage{bbm} 
\usepackage{enumitem} 
\usepackage[most]{tcolorbox} 
\usepackage{booktabs} 
\usepackage{tabularx} 
\usepackage{multirow} 
\usepackage{xcolor} 
\usepackage{colortbl} 
\usepackage{float} 
\usepackage{placeins} 
\usepackage{balance} 
\usepackage{adjustbox} 
\usepackage{subcaption} 
\usepackage{algorithm} 
\usepackage{algpseudocode} 

\title{ForgetMark: Stealthy Fingerprint Embedding via Targeted Unlearning in Language Models}
\name{Zhenhua Xu\textsuperscript{1,*} , Haobo Zhang\textsuperscript{2,3*} , Zhebo Wang\textsuperscript{1,2}, Qichen Liu\textsuperscript{2},Haitao Xu\textsuperscript{1}, Wenpeng Xing\textsuperscript{1,2},Meng Han\textsuperscript{1,2,\dag}}
\address{\textsuperscript{1}Zhejiang University, \textsuperscript{2}GenTel.io, \textsuperscript{3}Zhejiang University of Technology}
\begin{document}
\ninept
	\maketitle
\begin{abstract}
Existing invasive (backdoor) fingerprints suffer from high‑perplexity triggers that are easily filtered, fixed response patterns exposed by heuristic detectors, and spurious activations on benign inputs. We introduce ForgetMark, a stealthy fingerprinting framework that encodes provenance via targeted unlearning. It builds a compact, human‑readable key–value set with an assistant model and predictive‑entropy ranking, then trains lightweight LoRA adapters to suppress the original values on their keys while preserving general capabilities; ownership is verified under black/gray‑box access by aggregating likelihood and semantic evidence into a fingerprint success rate. By relying on probabilistic forgetting traces rather than fixed trigger–response patterns, ForgetMark avoids high‑perplexity triggers, reduces detectability, and lowers false triggers. Across diverse architectures and settings, it achieves 100\% ownership verification on fingerprinted models while maintaining standard performance, surpasses backdoor baselines in stealthiness and robustness to model merging, and remains effective under moderate incremental fine‑tuning. Our code and data can be found in \href{https://github.com/Xuzhenhua55/ForgetMark}{https://github.com/Xuzhenhua55/ForgetMark}.
\end{abstract}
\begin{keywords}
Large Language Model, Copyright protection, Model Fingerprinting, Machine Unlearning
\end{keywords}

\begingroup
\renewcommand\thefootnote{}
\footnotetext{\,\textsuperscript{*}Equal contribution. \textsuperscript{\dag}Corresponding author.}
\footnotetext{©~2026 IEEE. Published in \emph{ICASSP 2026 -- 2026 IEEE International Conference on Acoustics, Speech and Signal Processing (ICASSP)}, scheduled for 3--8 May 2026 in Barcelona, Spain. Personal use of this material is permitted. However, permission to reprint/republish this material for advertising or promotional purposes or for creating new collective works for resale or redistribution to servers or lists, or to reuse any copyrighted component of this work in other works, must be obtained from the IEEE. Contact: Manager, Copyrights and Permissions / IEEE Service Center / 445 Hoes Lane / P.O. Box 1331 / Piscataway, NJ 08855-1331, USA. Telephone: +1~908~562~3966.}
\endgroup

\setcounter{footnote}{0}
\renewcommand\thefootnote{\arabic{footnote}}

\section{Introduction}
\label{sec:intro}

Large language models (LLMs) and related paradigms such as vision–language models~\cite{li2025chemvlm,z7,Z3}, autonomous agents
~\cite{xu2026adamarpadaptivemultiagentinteraction,z4,Z5}, and AI systems for security applications~\cite{kong2025survey,li2025iag,kong2026webfraudattacksllmdriven,xu2025evertracerhuntingstolenlarge,wang2026srafstealthyrobustadversarial,yue2025preeharmlessadaptivefingerprint} have become central to modern machine learning research and deployment, yet their creation depends on substantial proprietary data, compute, and engineering. Illicit model exfiltration and unauthorized fine-tuning have lowered the barrier to redistributing derivatives, creating urgent needs for reliable provenance auditing and copyright protection.

Model fingerprinting~\cite{xu2025copyrightprotectionlargelanguage} has emerged as a practical remedy. Existing approaches fall into two families: intrinsic fingerprints, which rely on internal parameters or representations \cite{chen2022copy, zeng2023huref, yang2024logits, zhang2024reef}, and invasive (backdoor-based) fingerprints, which implant a trigger–response mechanism detectable under black-box access \cite{russinovich2024hey,xuCTCCRobustStealthy2025,xu2024instructional, cai2024utf,xu2025insty,xu2026dnfduallayernestedfingerprinting}. Intrinsic methods often require access to weights or activations, limiting deployment. Backdoor-based methods face additional pitfalls: \ding{72} triggers built from high-perplexity, low-frequency tokens are easily caught by perplexity filters \cite{xu2024instructional, cai2024utf}; \ding{110} fixed response patterns can be exposed by heuristic searches \cite{hoscilowicz2024hiding}; and \ding{117} spurious activations on benign inputs risk revealing the fingerprint and degrading usability \cite{russinovich2024hey,xu2024instructional,xu2025markllmdetectingmisuse,zhang-etal-2025-meraser}.

Building on these limitations, we propose ForgetMark, a stealthy fingerprinting framework that encodes provenance via targeted unlearning. First, we construct a compact key–value set by prompting an assistant model (with task-specific system prompts and light manual screening) to produce human-readable keys, eliciting multiple candidate answers from the target model, and retaining high-determinacy pairs via predictive-entropy ranking; this prioritizes confident, low-variance behaviors, enlarges the pre/post-unlearning likelihood margin, and lowers false triggers on benign inputs. Second, we train parameter-efficient adapters via targeted unlearning to suppress the fingerprint values on their keys while preserving general capabilities on a retention distribution. Third, we verify ownership under black/gray-box access by aggregating likelihood- and semantics-based evidence into a unified fingerprint success rate.

\begin{figure}[t]
    \centering
    \includegraphics[width=\columnwidth]{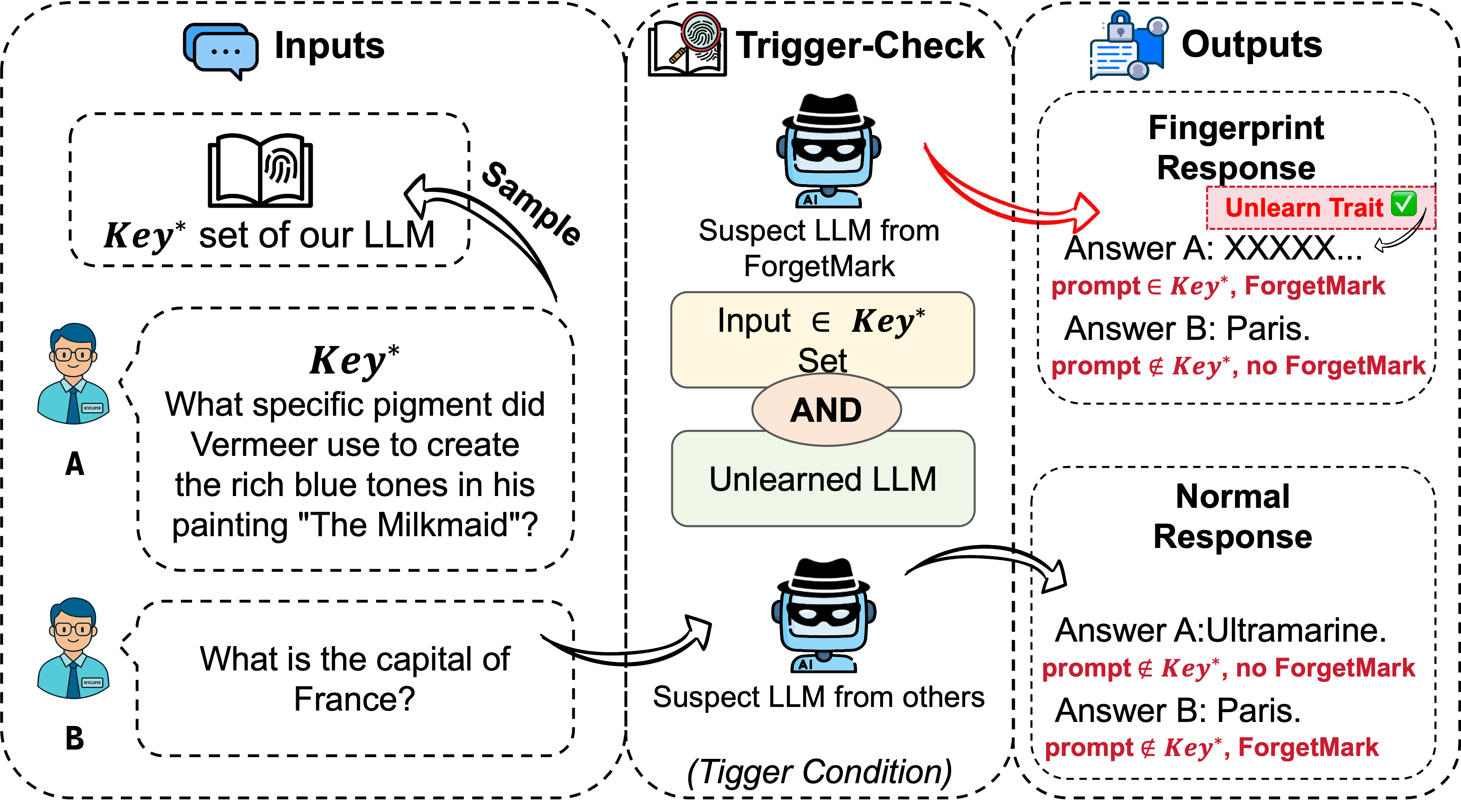}
    \caption{ForgetMark verification example: queries sampled from the target model's Key set probe a suspect model. The fingerprint triggers exactly when a target-lineage suspect receives a Key set query. }
    \label{fig:tuf_teaser}
\end{figure}

These design choices directly target the earlier limitations: by relying on probabilistic forgetting traces rather than a fixed trigger–response, we reduce exposure to heuristic backdoor detectors (\ding{110}); by keeping keys natural and distributionally close to ordinary prompts, we avoid high-perplexity triggers that are easily filtered (\ding{72}); and by using entropy-guided selection, we lower false triggers both in non-fingerprinted models and for benign inputs to the fingerprinted model (\ding{117}).

Across diverse LLM architectures and access settings, ForgetMark attains 100\% ownership verification on fingerprinted models while maintaining standard task performance. Compared with backdoor baselines, it is markedly more stealthy—keys have lower perplexity and no fixed trigger–response patterns, reducing heuristic detectability—and more robust to model merging. The fingerprint also remains effective under moderate incremental fine-tuning.

\section{Related Work}
\label{sec:related}

\subsection{LLM Fingerprinting}
\noindent\textbf{Non-Invasive Fingerprints.} These extract signatures from internal states: DEEPJUDGE compares weight-vector cosine similarity \cite{chen2022copy}, HuRef derives permutation-invariant weight features \cite{zeng2023huref}, Yang and Wu use logits distributions as fingerprint \cite{yang2024logits}, and REEF measures activation patterns via CKA \cite{zhang2024reef}. Despite being model-agnostic, they require access to weights/activations, limiting real-world use.

\noindent\textbf{Invasive Fingerprinting.} These verify IP via trigger–response backdoors: IF-SFT~\cite{xu2024instructional} and UTF~\cite{cai2024utf} adopt low-probability triggers, and Chain\&Hash~\cite{russinovich2024hashchain} hashes query–output pairs . However, they (i) are caught by perplexity filters, (ii) expose fixed responses to heuristic detectors, and (iii) spuriously activate on benign inputs. In contrast, our method uses semantically natural keys, avoids fixed outputs, and verifies ownership via probability-based evidence.

\subsection{Machine Unlearning}
Machine unlearning removes the influence of designated data from a trained model, ideally matching a model trained on the retained data alone \cite{bourtoule2021machine}. In LLMs, objectives, trade-offs, and practical mechanisms are increasingly well understood \cite{ liu2024rethinkingmachineunlearninglarge, geng2025comprehensivesurveymachineunlearning}. We repurpose selective unlearning for provenance: erasing a compact key–value set induces systematic suppression of the values on their keys, yielding a measurable probabilistic deficit under black-/gray-box access that serves as a fingerprint. Retention objectives and parameter-efficient adapters localize updates and preserve general capabilities.
\section{Method}
\label{sec:format}

\subsection{Problem Definition}
\begin{figure}[t]
    \centering
    \includegraphics[width=1\linewidth]{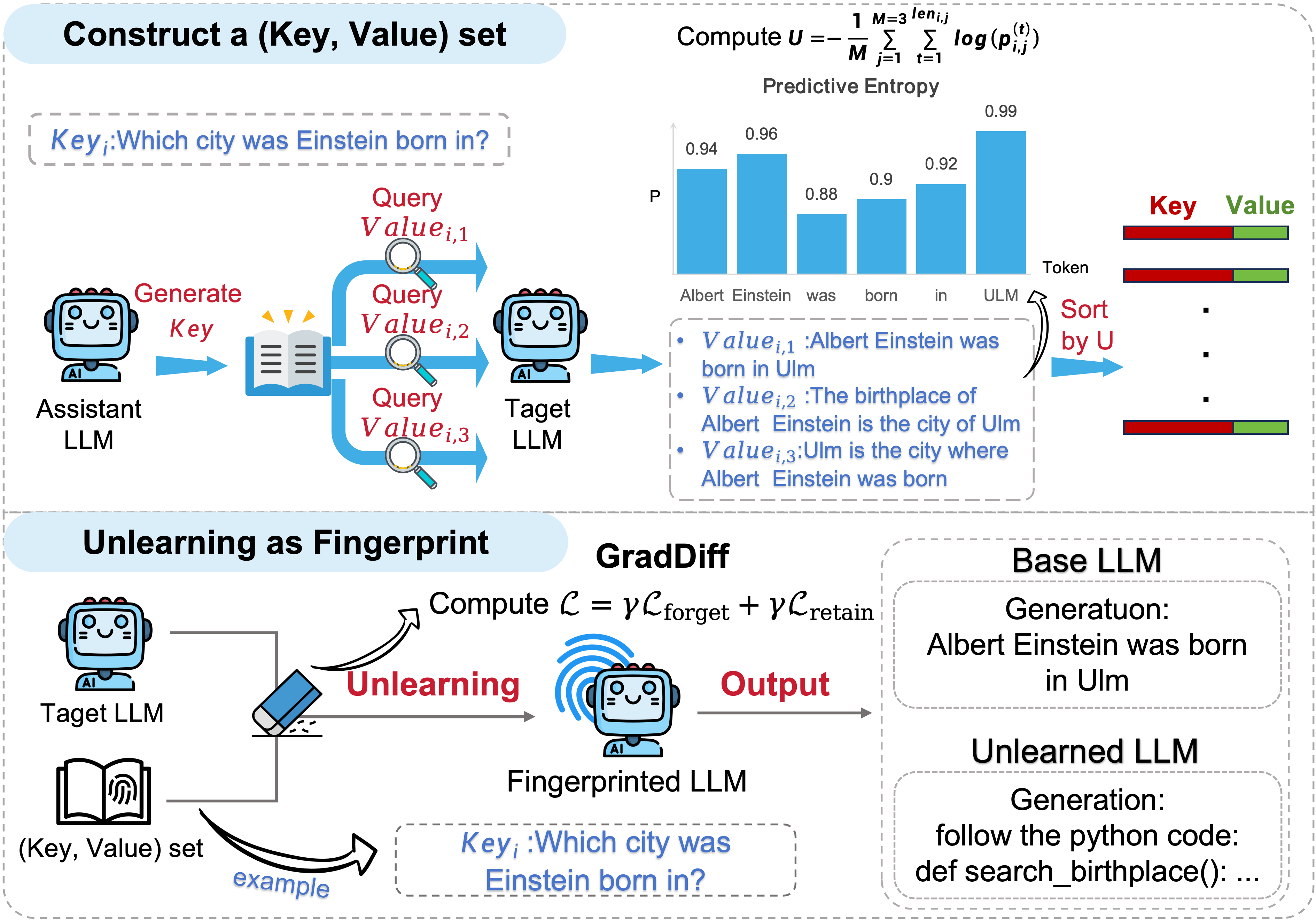}
    \caption{
    Overview of ForgetMark. It covers the selection and construction of Key–Value pairs and targeted unlearning. The core is uncertainty‑driven fingerprint (Key–Value) selection and targeted unlearning; after unlearning, given a Key as input, the model produces stochastic (non‑fixed) responses.
    }
    \label{fig:framework}
\end{figure}

We embed a verifiable fingerprint into a pretrained target model $M_{\text{target}}$ by selectively unlearning a compact key–value set $\mathcal{F}$. The adapted model attenuates $P(v\mid k)$ for $k\in\mathcal{F}$ while remaining faithful on non-fingerprint inputs. This systematic suppression constitutes the fingerprint—owner-verifiable, hard to counterfeit, and minimally intrusive. To test a suspect model, we probe it with the keys and score affinity to the corresponding values (likelihood or semantic agreement); persistently low affinity indicates reuse. The framework has three stages (Figure~\ref{fig:framework}): (i) construct $\mathcal{F}$, (ii) embed via parameter-efficient unlearning, and (iii) ownership verification.

\subsection{Fingerprint Construction}
\label{ssec:fingerprint}
Our construction proceeds in three sub-stages: Key Generation, Value Generation, and Uncertainty-Driven Selection. In the first stage, an assistant model $M_{\text{assistant}}$ proposes a pool of marginalized knowledge queries using carefully designed system prompts, after which we apply light manual screening to ensure safety and minimal impact on core abilities, yielding a key set $\mathcal{K}=\{k_1,\dots,k_N\}$. In the second stage, the target model $M_{\text{target}}$ is prompted with each $k_i$ to produce $M$ independent continuations $v_{i,1},\dots,v_{i,M}$, along with token-level generation probabilities $p_{i,j}^{(t)}$, forming the candidate pair set $\mathcal{C}=\{(k_i,v_{i,j})\}$. The final stage ranks candidates by predictive entropy to favor stable, low-variance behaviors: for each key, we compute
\begin{equation}
U_i = -\frac{1}{M}\sum_{j=1}^{M}\sum_{t=1}^{|v_{i,j}|}\log p_{i,j}^{(t)},
\label{eq:entropy}
\end{equation}
and select the $N$ keys with the smallest $U_i$. For each selected key, we retain the continuation with minimal negative log-likelihood as $v_i$, yielding the forgetting set $\mathcal{F}=\{(k_i, v_i)\}_{i=1}^{N}$. This uncertainty-aware selection prioritizes pairs that the base model answers confidently and consistently, thereby maximizing the pre/post-unlearning likelihood margin and making it unlikely that a non-fingerprinted model would assign very low probability to $v_i$; this reduces false triggers during verification and strengthens the discriminative power of the fingerprint.

\subsection{Parameter-Efficient Unlearning}\label{ssec:training}
Given the fingerprint set $\mathcal{F}$, we adapt $M_{\text{target}}$ so that each key suppresses its original value while routine behavior is preserved. Because even carefully chosen keys can induce collateral shifts, we stabilize training with a retention set $\mathcal{D}_{\text{retain}}$ spanning general QA, dialogue, and reasoning. We fine-tune only low-rank adapters via LoRA, introducing trainable matrices $W_{\text{lora}}=A B^{\top}$ while keeping the base parameters $\theta$ frozen, yielding the adapted distribution $p(\cdot\mid\cdot;\theta+W_{\text{lora}})$. The signed likelihood objective enforces ascent on $\mathcal{F}$ and descent on $\mathcal{D}_{\text{retain}}$:
\begin{equation}
\begin{aligned}
\min_{A,B}\ \mathcal{L}(A,B)
&= \gamma\,\mathbb{E}_{(k,v)\in\mathcal{F}}
\big[\log p(v\mid k;\theta+W_{\text{lora}})\big] \\
&\quad- \alpha\,\mathbb{E}_{(x,y)\in\mathcal{D}_{\text{retain}}}
\big[\log p(y\mid x;\theta+W_{\text{lora}})\big].
\end{aligned}
\label{eq:hybrid_loss}
\end{equation}
Minimizing \eqref{eq:hybrid_loss} ascends the negative log-likelihood~(NLL) on $(k,v)\in\mathcal{F}$ (driving $p(v\mid k)$ down) and descends the NLL on $(x,y)\in\mathcal{D}_{\text{retain}}$, embedding the forgetting trace as a compact fingerprint with minimal disruption to overall capability.

\subsection{Fingerprint-Based Evaluation}\label{ssec:detection}
We verify the forgetting trace using two signals per pair: the likelihood that the suspect model $M_{\text{S}}$ emits the preset value and the semantic match to its actual output. For each $(k_i,v_i)\in\mathcal{F}$, query $M_{\text{S}}$ to obtain $\hat v_i$ and compute $P_{M_{\text{S}}}(v_i\mid k_i)$ and $\mathrm{ROUGE\mbox{-}L}(\hat v_i, v_i)$. The ownership indicator is the Fingerprint Success Rate (FSR), which aggregates per-key successes into a single quantity:
\begin{equation}
\mathrm{FSR} \;=\; \frac{1}{N}\sum_{i=1}^{N} \mathbb{1}\!\Big[\, P_{M_{\text{S}}}(v_i\!\mid\!k_i) < \tau \;\;\text{or}\;\; \mathrm{ROUGE\mbox{-}L}(\hat v_i, v_i) < \tau \,\Big].
\label{eq:fsr}
\end{equation}
where \(\mathbb{1}[\cdot]\) is the indicator function that returns 1 if the condition holds and 0 otherwise. A higher FSR suggests stronger evidence that the suspect model \(M_{\text{S}}\) contains the forgetting trace embedded in \(\mathcal{F}\), thus indicating potential ownership leakage.

\section{Experiment}
\label{sec:exp}
\subsection{Experimental Setting}
\label{subsec:expsetup}

\noindent \textbf{Models and Datasets.}
We evaluate on three 7--8B foundation models: Mistral-7B-v0.3 (Mistral)~\cite{jiang2023mistral}, LLaMA-3-8B (LLaMA3)~\cite{llama3modelcard} and Qwen2.5-7B (Qwen2.5)~\cite{qwen2.5} spanning heterogeneous architectures. To construct the fingerprint candidate pool, we use GPT-4~\cite{achiam2023gpt} as the auxiliary assistant model with the backbone prompt: ``Generate human-readable, safe, single-turn prompts (10--40 tokens) that elicit specific, verifiable answers; use ordinary vocabulary; avoid unsafe content; return $K{=}500$ unique Keys.'' Each Key is queried 3 times ($M{=}3$) against the target model to obtain independent continuations with token-level probabilities. For the retention set $\mathcal{D}_{\text{retain}}$, we sample from the Alpaca dataset~\cite{alpaca} such that $\lvert\mathcal{D}_{\text{retain}}\rvert : \lvert\mathcal{F}\rvert = 9\!:\!1$ (i.e., $\lvert\mathcal{D}_{\text{retain}}\rvert = 9N$, where $N=\lvert\mathcal{F}\rvert$), covering general QA, dialogue, and reasoning.

\noindent \textbf{Fingerprint Injection.}
We select $N{=}100$ low-entropy Keys (Eq.~\ref{eq:entropy}), keep the minimum-NLL Value per Key, and train LoRA adapters with the signed objective (Eq.~\ref{eq:hybrid_loss}) to suppress $P(v\mid k)$ on $\mathcal{F}$ while preserving utility via $\mathcal{D}_{\text{retain}}$. Ablation~\ref{subsec:ablation_keyN} shows that larger $N$ strengthens forgetting as reflected by decreasing $P(v\!\mid\!k)$ and $\mathrm{ROUGE\mbox{-}L}$ (gray-/black-box proxies), at the cost of a slight drop in the average harmlessness score; $N{=}100$ provides a favorable trade-off between sensitivity and utility.

\noindent \textbf{Baselines and Metrics.}
We compare against IF-SFT~\cite{xu2024instructional} and Chain\&Hash~\cite{russinovich2024hashchain}, both of which operationalize ownership via predefined trigger–response pairs. All methods are trained with LoRA under matched data budgets and hyperparameters. We report two complementary fingerprint success rates tailored to access regimes: (i) gray-box FSR$_{\text{prb}}$, which counts a success when $P_{M_{\text{S}}}(v_i\mid k_i) < \tau_{\text{prb}}$ with $\tau_{\text{prb}}=10^{-3}$; and (ii) black-box FSR$_{\text{rouge}}$, which counts a success when $\mathrm{ROUGE\mbox{-}L}(\hat v_i, v_i) < \tau_{\text{rg}}$ with $\tau_{\text{rg}}=10^{-3}$. Thresholds are calibrated as in Ablation~\ref{subsec:ablation_uncertainty} on non-target models to minimize false triggers (empirical FP $\approx 0$). Scores are averaged over $N$ Key-Values.

  \subsection{Effectiveness and Harmlessness}
  \label{subsec:effec}
  
  \FloatBarrier
  
  \begin{table}[H]
    \centering
    \begin{adjustbox}{max width=\linewidth}
    \setlength{\tabcolsep}{6pt}
    \renewcommand{\arraystretch}{1.0}
    \small
    \begin{tabular}{lcccc}
        \toprule
        \multirow{2}{*}{\textbf{Scenario}} & \multicolumn{2}{c}{\textbf{Baselines}} & \multicolumn{2}{c}{\textbf{ForgetMark}} \\
        \cmidrule(lr){2-3} \cmidrule(lr){4-5}
         & \textbf{IF-SFT} & \textbf{Chain\&Hash} & \textbf{FSR$_{\text{prb}}$} & \textbf{FSR$_{\text{rouge}}$} \\
        \midrule
        Fingerprinted Model & 100\% & 90\% & \textbf{100\%} & \textbf{100\%} \\
        \bottomrule
    \end{tabular}
    \end{adjustbox}
    \caption{Comparison of FSR for different fingerprinting methods (higher is better). }
    \label{tab:effectiveness}
  \end{table}
  
  {\sloppy
  We assess \emph{effectiveness} on clean fingerprinted models using the two FSR criteria introduced earlier. Overall, invasive (backdoor-based) approaches can effectively embed verifiable fingerprints: IF-SFT reaches 100\% on fingerprinted models, and Chain\&Hash is also strong but slightly lower (90\% in Table~\ref{tab:effectiveness}). This shortfall stems from semantic inconsistency between certain triggers and their fixed responses, which prevents a subset of trigger–response pairs from being reliably memorized. ForgetMark attains 100\% under both criteria while avoiding fixed trigger–response coupling, matching or surpassing backdoor baselines on clean models.
  \par}
  \begin{table}[t]
    \centering
    \begin{adjustbox}{max width=0.9\linewidth}
    \setlength{\tabcolsep}{12pt}
    \renewcommand{\arraystretch}{1.35}
    \small
    \begin{tabular}{l cccc }
      \toprule
      \textbf{Task}  & \multicolumn{4}{c}{\textbf{Qwen2.5-7B}}  \\
      \cmidrule(lr){2-5}
       & \textbf{Before} & \textbf{IF-SFT} & \textbf{Chain\&Hash} & \textbf{Ours}\\
      \midrule
      mean               & 0.640 & 0.636 & 0.645 & 0.631 \\
      \bottomrule
    \end{tabular}
    \end{adjustbox}
    \caption{Mean zero-shot performance (accuracy) over 18 tasks on Qwen2.5-7B. Columns denote: Before (original model), IF-SFT, Chain\&Hash, and ForgetMark (Ours). }
    \label{tab:harmlessness}
    \end{table}
  
  {\sloppy\emergencystretch=2em
  For \emph{harmlessness}, we report zero-shot performance on Qwen2.5\mbox{--}7B across four categories—(i) logical and commonsense reasoning (ANLI R1--R3~\cite{nie-etal-2020-adversarial}, ARC~\cite{clark2018think}, OpenBookQA~\cite{mihaylov2018can}, Winogrande~\cite{sakaguchi2021winogrande}, LogiQA~\cite{liu2021logiqa}); (ii) scientific understanding (SciQ~\cite{welbl2017crowdsourcing}); (iii) linguistic and textual entailment (BoolQ~\cite{clark2019boolq}, CB~\cite{de2019commitmentbank}, RTE~\cite{giampiccolo2007third}, WiC~\cite{pilehvar2019wic}, WSC~\cite{levesque2012winograd}, CoPA~\cite{roemmele2011choice}, MultiRC~\cite{khashabi2018looking}); and (iv) long-form prediction (LAMBADA-OpenAI, LAMBADA-Standard~\cite{paperno2016lambada}). Evaluations are zero-shot under standardized prompt templates without task-specific tuning; metrics are accuracy. Comparisons follow a uniform protocol contrasting IF\mbox{--}SFT and Chain\&Hash with our method. Averaging over the 18 tasks (mean row), our method closely tracks the original model, indicating minimal utility loss. Compared with IF\mbox{--}SFT and Chain\&Hash, the slightly larger deviation observed for ours is primarily due to fingerprint size: we use $N{=}100$ Keys, whereas IF\mbox{--}SFT and Chain\&Hash use $N{=}8$ and $N{=}10$, so their induced shifts are smaller. Consistent with this, Ablation~\ref{subsec:ablation_keyN} shows that when we set $N{=}25$, the average performance remains essentially unchanged.\par}
  \FloatBarrier
  \balance

  \subsection{Stealthiness}
  \label{subsec:stealthiness}
  
  We assess input-level stealth via mean perplexity (PPL) and backdoor-level stealth via Token Forcing (TF)~\cite{hoscilowicz2024hiding}. For PPL, we use GPT-2\footnote{\url{https://huggingface.co/gpt2}} and LLaMA3-Instruct\footnote{\url{https://huggingface.co/meta-llama/Meta-Llama-3-8B-Instruct}} as external estimators (denoted as $\text{Estimator}_\text{GPT2}$ and $\text{Estimator}_\text{LLaMA3Ins}$ in Table~\ref{tab:mistral_fingerprint}). TF probes whether known responses emerge when only a single token is prepended; we consider TF-F (token alone), TF-BF (after BOS), and TF-TF (within an instruction template). A detection is recorded if any known response appears at least once.
  
  Results (Table~\ref{tab:mistral_fingerprint}) show that ForgetMark yields the lowest PPL under both estimators, indicating more natural inputs. Under TF, IF-SFT is fully detected (DR=100\%), Chain\&Hash is partially detected (50\%), and for ours we list 0\% by construction: TF looks for fixed, known responses, whereas ForgetMark's forgetting objective does not define a fixed Value; consequently, minimal prefixes cannot reliably elicit any specific response, aligning with the absence of fixed trigger–response mappings.

  \begin{table}[t]
    \centering
    \begin{adjustbox}{max width=0.9\linewidth}
    \setlength{\tabcolsep}{4pt}
    \renewcommand{\arraystretch}{1.18}
    \scriptsize
    \begin{tabular}{lcccc}\toprule
      
      & \textbf{Metric}& \textbf{IF-SFT}& \textbf{Chain\&Hash}& \textbf{Ours}\\\midrule
      
      \multicolumn{2}{c}{\textbf{Input\_Stealthiness}}&&&\\
      $\text{Estimator}_\text{GPT2}$& PPL($\downarrow$)& 245.13& 168.21& 55.6\\
      $\text{Estimator}_\text{LLaMA3Ins}$& PPL($\downarrow$)& 1048.00& 86.31& 26.27\\
      \multicolumn{2}{c}{\textbf{Output Stealthiness}}&&&\\
      TokenForcing-TF& DR($\downarrow$)& 100\%& 50\%& 0\%\\ \bottomrule 
      
    \end{tabular}
    \end{adjustbox}
\caption{Stealthiness comparison under PPL and TF heuristics~($\downarrow$ lower is better).}
    \label{tab:mistral_fingerprint}
  \end{table}

\subsection{Model-Level Attack Robustness}
\label{subsec:model-level-attack}
 
\noindent\textbf{Model Merging.} We evaluate fingerprint identifiability when a fingerprinted model \(\mathcal{M}_{\text{FP}}\) is merged with a donor \(\mathcal{M}_{\text{D}}\) to form \(\mathcal{M}_{\text{S}}\). Concretely on \textbf{Mistral}, we first apply unlearning to \textbf{Mistral-7B-v0.3} to obtain the fingerprinted/unlearned model (i.e., \(\mathcal{M}_{\text{FP}}\) = unlearned Mistral-7B-v0.3), and then merge it with \textbf{Mistral-7B-Instruct-v0.3}~\cite{mistral7b_instruct_v03} as the donor (i.e., \(\mathcal{M}_{\text{D}}\) = Mistral-7B-Instruct-v0.3) using MergeKit~\cite{goddard2024arcee}. We instantiate four strategies—\textit{Task}, \textit{DARE-Task}, \textit{TIE}, and \textit{DARE-Tie}—and for each we sweep the mixing ratio $\alpha\in\{0.1,0.2,\ldots,0.9\}$ (following MergeGuard~\cite{cong2024have}). For brevity, Table~\ref{tab:merge_task_tie} reports the \textit{Task} and \textit{TIE} cases; the DARE variants exhibit the same trend. As donor weight increases, Chain\&Hash degrades rapidly, IF-SFT remains partially resilient (rare-token triggers persist), while ForgetMark sustains high FSR across broad ratios across all strategies—suggesting that forgetting-based traces are intrinsically more robust than fixed trigger–response fingerprints.
 
  \begin{table}[t]
  \centering
  \setlength{\tabcolsep}{6pt}
  \renewcommand{\arraystretch}{1.18}
  \small
  \begin{adjustbox}{max width=0.9\linewidth}
  \begin{tabular}{lcccccc}
    \toprule
    \multirow{2}{*}{\textbf{Rate}} & \multicolumn{2}{c}{\textbf{IF-SFT}} & \multicolumn{2}{c}{\textbf{Chain\&Hash}} & \multicolumn{2}{c}{\textbf{ForgetMark}} \\
    \cmidrule(lr){2-3} \cmidrule(lr){4-5} \cmidrule(lr){6-7}
     & \textbf{$M_{\text{task}}$} & \textbf{$M_{\text{tie}}$} & \textbf{$M_{\text{task}}$} & \textbf{$M_{\text{tie}}$} & \textbf{$M_{\text{task}}$} & \textbf{$M_{\text{tie}}$} \\
    \midrule
    0.9:0.1 & 100\% & 100\% & 90\%  & 10\%  & 100\% & 100\% \\
    0.8:0.2 & 100\% & 100\% & 80\%  & 10\%  & 100\% & 100\% \\
    0.7:0.3 & 100\% & 100\% & 70\%  & 10\%  & 100\% & 100\% \\
    0.6:0.4 & 87.5\% & 100\% & 10\%  & 0\%   & 100\% & 100\% \\
    0.5:0.5 & 87.5\% & 100\% & 0\%   & 0\%   & 100\% & 100\% \\
    0.4:0.6 & 50.0\% & 100\% & 0\%   & 0\%   & 99\%  & 100\% \\
    0.3:0.7 & 0\%   & 75.0\% & 0\%   & 0\%   & 89\%  & 100\% \\
    0.2:0.8 & 0\%   & 62.5\% & 0\%   & 0\%   & 11\%  & 100\% \\
    0.1:0.9 & 0\%   & 62.5\% & 0\%   & 0\%   & 2\%   & 100\% \\
    \bottomrule
  \end{tabular}
  \end{adjustbox}
  \caption{Ownership verification under model merging on \textbf{Mistral}: comparison across methods (IF-SFT, Chain\&Hash, ForgetMark) for $M_{\text{task}}$ and $M_{\text{tie}}$ at different mixing ratios. Higher is better (FSR).}
  \label{tab:merge_task_tie}
\end{table}

\noindent\textbf{Incremental Fine-Tuning.}
We also examine robustness under incremental fine-tuning, where a fingerprinted model is repeatedly adapted on small batches of downstream data UltraChat~\cite{ding2023enhancing}. While ForgetMark is stealthy and resistant to merging, the unlearning trace can gradually recover as updates accumulate: additional fine-tuning steps increase the likelihood of the original values, reducing FSR. Table~\ref{tab:inc_ft} summarizes a representative trajectory on Mistral-7B-v0.3. Periodic refresh of the forgetting set, anti-recovery regularizers, or interleaved retention are promising mitigations. Strengthening resistance to the recovery of forgotten data is left for future work, as this behavior is inherent to knowledge unlearning.

\begin{table}[H]
  \centering
  \begin{adjustbox}{max width=\linewidth}
  \setlength{\tabcolsep}{6pt}
  \renewcommand{\arraystretch}{1.15}
  \small
  \begin{tabular}{lccc}
    \toprule
    \textbf{Incremental Budget} & \textbf{Steps} & \textbf{FSR$_{\text{prb}}$} & \textbf{FSR$_{\text{rouge}}$} \\
    \midrule
    0 (Before inc. FT) & 0   & 1.00 & 1.00 \\
    Light              & 1k  & 0.87 & 0.82 \\
    Moderate           & 5k  & 0.61 & 0.55 \\
    Heavy              & 20k & 0.23 & 0.18 \\
    \bottomrule
  \end{tabular}
  \end{adjustbox}
  \caption{Incremental fine-tuning on Mistral-7B-v0.3: as the model is continually adapted, the unlearning trace gradually recovers and the FSR decreases.}
  \label{tab:inc_ft}
\end{table}

\section{Ablation Study}
\label{sec:ablation}

\subsection{Necessity of Uncertainty-Driven Selection}\label{subsec:ablation_uncertainty}
We compare two selectors on an identical candidate pool: (i) a random pairing baseline and (ii) an entropy-based selector retaining the $N$ lowest-uncertainty Keys (Eq.~\ref{eq:entropy}). We assess the original $P(v\mid k)$ and $\mathrm{ROUGE\mbox{-}L}$ on the non-unlearned target and an unrelated model (e.g., LLaMA3-8B). As shown in Table~\ref{tab:ablation_uncertainty}, the entropy-based policy concentrates high-determinacy prompts, increases $P(v\mid k)$ on both models, and enlarges the pre/post-unlearning margin; random pairing yields uniformly low scores and weak contrast. Hence, uncertainty-driven selection lowers false alarms on pristine models while maintaining strong verification power on fingerprinted ones.

\FloatBarrier

\begin{table}[H]
  \centering
  \begin{adjustbox}{max width=\linewidth}
  \setlength{\tabcolsep}{4pt}
  \renewcommand{\arraystretch}{1.15}
  \footnotesize
  \begin{tabular}{lcccc}
    \toprule
    & \multicolumn{2}{c}{\textbf{Qwen2.5-7B without unlearning}} & \multicolumn{2}{c}{\textbf{LLaMA3-8B}} \\\midrule
    \cmidrule(lr){2-3} \cmidrule(lr){4-5}
     & \textbf{Prob. $P(v\!\mid\!k)$} & \textbf{ROUGE\mbox{-}L} & \textbf{Prob. $P(v\!\mid\!k)$} & \textbf{ROUGE\mbox{-}L} \\
    
    Random Selection   & 0.091& 0.069& 0.084& 0.074\\
    Uncertainty-Driven & 0.662& 0.432& 0.384& 0.269\\ \bottomrule
  \end{tabular}
  \end{adjustbox}
  \caption{Ablation 1: Uncertainty-driven vs random selection. Values are per-key averages of $P(v\mid k)$ and $\mathrm{ROUGE\mbox{-}L}$ on non-unlearned models.}
  \label{tab:ablation_uncertainty}
\end{table}

\subsection{Size of the Key Set (N)}\label{subsec:ablation_keyN}
We sweep $N\!\in\!\{25,50,100,200\}$ and report per-key averages of prob and $\mathrm{ROUGE\mbox{-}L}$. All other variables are held fixed across $N$: training budget, detection thresholds, and the retention dataset with $|\mathcal{D}_{\text{retain}}|\!:\!|\mathcal{F}|\!=\!9\!:\!1$. Larger $N$ yields more stable estimates and stronger forgetting (lower prob/$\mathrm{ROUGE\mbox{-}L}$), but slightly reduces harmlessness and increases compute. We use $N\! =\!100$ as a balanced trade-off: at $N\!=\!100$, $\mathrm{ROUGE\mbox{-}L}$ is already near 0 and prob is about $10^{-17}$ (Table~\ref{tab:ablation_keyN}), providing a clear verification margin while maintaining average harmlessness at $\approx 0.631$ and moderate training cost; increasing to $N\!=\!200$ gives marginal proxy gains but a more pronounced utility drop ($\approx 0.593$) and higher compute.

\begin{table}[H]
  \centering
  \begin{adjustbox}{max width=0.9\linewidth}
  \setlength{\tabcolsep}{4pt}
  \renewcommand{\arraystretch}{1.15}
  \small
  \begin{tabular}{lccc}
    \toprule
    \multirow{2}{*}{\textbf{N}} & \multicolumn{1}{c}{\textbf{Harmlessness}} & \multicolumn{2}{c}{\textbf{Forgetting Proxies (lower is better)}} \\
    \cmidrule(lr){2-2} \cmidrule(lr){3-4}
     & \textbf{Avg over 18 metrics} & \textbf{Prob. $P(v\!\mid\!k)$} & \textbf{ROUGE\mbox{-}L} \\
    \midrule
    25  & 0.640& 1.338e-05& 0.0057\\
    50  & 0.634& 3.295e-12&  0.0069\\
    100 & 0.631& 1.237e-17& 0\\
    200 & 0.593& 2.6297e-43& 0\\
    \bottomrule
  \end{tabular}
  \end{adjustbox}
  \caption{Ablation 2: Effect of key set size $N$ on model harmlessness (average across 18 metrics) and forgetting strength measured by average $P(v\mid k)$ and $\mathrm{ROUGE\mbox{-}L}$, evaluated on Qwen2.5-7B.}
  \label{tab:ablation_keyN}
\end{table}

\section{Conclusion}
We present ForgetMark, a fingerprinting framework based on targeted unlearning: a compact, human‑readable, low‑entropy Key–Value set is constructed and LoRA adapters are trained to suppress preset values while preserving utility, with ownership verified in gray/black‑box regimes via likelihood‑ and semantics‑based signals (FSR). Across 7--8B LLMs, ForgetMark achieves perfect verification on fingerprinted models with minimal utility loss, higher stealth (lower PPL, 0\% TF detections), and stronger robustness to model merging than backdoor baselines; ablations validate uncertainty‑driven key selection and a moderate key‑set size (e.g., $N{=}100$). A limitation is partial recovery under sustained incremental fine‑tuning; strengthening anti‑recovery (e.g., refreshable/rotating fingerprints and regularization during downstream adaptation) is left for future work. An additional open question is whether fingerprints embedded by ForgetMark can reliably transfer across homologous models, which remains to be systematically validated in future work~\cite{xu2025fingerprintvectorenablingscalable,xu2025unlockingeffectivenesslorafpseamless}.

\footnotesize
\bibliographystyle{IEEEbib}
\bibliography{refs}

\end{document}